\documentclass{DISproc}

\begin{document}
\title{Status on the transversity parton distribution: the dihadron fragmentation functions way}

\author{{\slshape A.~Courtoy$^1$, Alessandro~Bacchetta$^{2, 3}$, Marco~Radici$^2$}\\[1ex]
$^1$IFPA, AGO Department,  Universit\'e de Li\`ege, All\'ee du 6 Ao\^ut 17, 4000 Li\`ege, Belgium\\
$^2$INFN Sezione di Pavia, via Bassi 6, I-27100 Pavia, Italy\\
$^3$Dipartimento di Fisica, Universit\`a di Pavia, via Bassi 6, I-27100 Pavia, Italy
}

\contribID{xy}

\doi  

\maketitle

\begin{abstract}
  We report on the extraction of dihadron fragmentation functions (DiFF) from the semi-inclusive production of two hadron pairs in back-to-back jets in $e^+e^-$ annihilation. A nonzero asymmetry in the correlation of azimuthal orientations of opposite $\pi^+\pi^-$ pairs is related to the transverse polarization of fragmenting quarks through a significant polarized DiFF. A combined analysis of this asymmetry and the spin asymmetry in the SIDIS process $ep^{\uparrow}\to e'(\pi^+\pi^-)X$ has led to the first extraction of the $u$ and $d$-flavor transversity parton distribution function in the framework of collinear factorization. 
  \end{abstract}


The distribution of quarks and gluons inside hadrons can be described by means of
parton distribution functions (PDFs). In a parton-model picture, PDFs describe
combinations of number densities of quarks and gluons in a fast-moving
hadron. The knowledge of PDFs is crucial for our understanding of QCD and for the interpretation of high-energy 
experiments involving hadrons. 
At leading-twist, the quark structure of hadrons is described by three PDFs. The unpolarized distribution $f_1(x)$, the longitudinal polarization, {\it helicity}, distribution $g_1(x)$ and the transverse polarization, {\it transversity}, distribution $h_1(x)$. 
From the phenomenological point of view, the unpolarized PDF is a well-known quantity and  the helicity PDF is known to some extent. On the other hand, the transversity distribution is poorly known. It is due to its chiral-odd nature. In effect, transversity  is not observable from fully inclusive DIS. In order to measure the transversity PDF, chirality must be flipped twice. We can either have two hadrons in the initial state, e.g. proton-proton collision ; or one hadron in the initial state and at least one hadron in the final state, e.g. semi-inclusive DIS (SIDIS).

In these proceedings, we consider two-hadron production in DIS, i.e., the process
\begin{equation}
\ell(l) + N(P) \to \ell(l') + H_1(P_1)+H_2(P_2) + X \enspace,
\label{eq:sidis}
\end{equation}
where $\ell$ denotes the beam lepton, $N$ the nucleon target, $H_1$ and $H_2$
the produced hadrons,  
and where four-momenta are given in parentheses. The transversity distribution function is here multiplied by a
chiral-odd DiFF, denoted as
$H_1^{\sphericalangle\,q}$~\cite{Radici:2001na}, which describes the correlation between the 
transverse polarization of the fragmenting quark with flavor $q$ and the azimuthal 
orientation of the plane containing the momenta of the detected hadron pair. DiFFs are functions of the invariant mass of the pion pair, $M_h$, and the momentum fraction of the fragmenting quark carried by the pair, $z$.
Since DiFFs are not TMD FF but collinear FF instead, this effect survives after integration over quark transverse 
momenta and can therefore be analyzed in the framework of collinear factorization.  
Measurement of the relevant asymmetry has been presented by
the HERMES collaboration for the production of $\pi^+ \pi^-$ pairs on transversely polarized 
protons~\cite{Airapetian:2008sk}, and led to the first extraction of transversity in a collinear framework~\cite{Bacchetta:2011ip}.  The COMPASS collaboration~\cite{Adolph:2012nw} has recently published similar measurements on proton and deuteron targets. The combination of these new data allows for a flavor separation of the transversity, which we present here for the first time. 
\\


In order to extract transversity for pion pair production in SIDIS, one has to determine independently the DiFFs. This can be achieved by studying the correlations between the azimuthal orientations of two pion pairs in back-to-back 
jets in $e^+e^-$ annihilation~\cite{Artru:1996zu}. %
The first analysis of the so-called Artru--Collins asymmetry~\cite{Boer:2003ya}, i.e.,
\begin{eqnarray} 
A_{e^+e^-}( z,  M_h, \overline{z},  \overline{M}_h ; Q^2) 
&\propto &
 \frac{\sum_q e_q^2 \, \frac{|{\bf R}|}{M_h} \,H_{1}^{\sphericalangle q}(z, M_h; Q^2)\;
         \frac{|\overline{{\bf R}}|}{\overline{M}_h} \, \overline{H}_{1}^{\sphericalangle q}(\overline{z}, \overline{M}_h; Q^2)}
      {\sum_q e_q^2\, D_1^q (z, M_h; Q^2) \; \overline{D}_1^q (\overline{z}, \overline{M}_h; Q^2)} \enspace,
\label{eq:ACasy}
\end{eqnarray} 
 by the Belle collaboration~\cite{Vossen:2011fk} made possible a direct extraction of $H_{1}^{\sphericalangle}$~\footnote{We call here  $H_{1,sp}^{\sphericalangle}$ by $H_{1}^{\sphericalangle}$ to avoid clumsy notations.} for the 
production of $\pi^+ \pi^-$.  
$R$ is the relative momentum of the pair, it obeys the relation
$
2\,|{\bf R}|/M_h = \sqrt{1-4m_\pi^2/M_h^2}.
$
In the absence of a measurement of the unpolarized cross section for dihadron
production in $e^+e^-$ annihilation, the unpolarized DiFF, $D_1$, 
was parametrized to reproduce the two-hadron yields of the PYTHIA event 
generator, which is known to give a good description of data. 
Combining the parametrization of the unpolarized functions $D_1$ with the fit
of the azimuthal asymmetry presented in Ref.~\cite{Vossen:2011fk}, 
it was possible to
 extract the DiFF $H_{1}^{\sphericalangle}$~\cite{Courtoy:2012ry}. 
 \\

\begin{wraptable}{r}{.53\textwidth}
  \centering
  \begin{tabular}{|rr|r|}
    \toprule
    &HERMES range & COMPASS range \\
        \midrule
    $n_u=$       			& $0.564$	 		   &	$0.785 $	\\
    $n_s=$       			& $0.303$	 		   &	$0.443 $	\\
    $n_u^{\uparrow}=$        & $0.146\pm  0.037$ 	   &	$0.163\pm  0.031$	\\
    \bottomrule
  \end{tabular}
  \caption{DiFF integrated over experimental ranges. We neglect the error coming from the determination of $D_1$.}
  \label{tab:nq_ratios}
\end{wraptable}

We have now all the ingredients at hand to extract transversity from the process~(\ref{eq:sidis}).
In a collinear framework, we can single out the DiFF contribution to the SIDIS asymmetry from the $x$-dependence coming from the transversity PDF. 
In particular, the $x$ behavior of $h_1(x)$ is simply given by integrating the numerator of the asymmetry over the $(z, M_h)$-dependence, so that the relevant quantities  for our purposes are
\\
\begin{eqnarray}
n_q(Q^2) &=& \int dz \, dM_h \, D_1^q (z, M_h; Q^2)  \enspace, \label{eq:nD1}\\
n_q^\uparrow (Q^2) &=& \int dz \, dM_h \, \frac{|{\bf R}|}{M_h}\, H_{1}^{\sphericalangle\, q}(z,M_h; Q^2) \enspace,
\label{eq:nH1angle}
\end{eqnarray}
where the integral limits are defined in the physical range of validity, i.e. from $2m_{\pi}$ to $M_h\ll Q$ and $0.2<z<1$. 

The asymmetry, expressed in terms of the integrated DiFFs, reads
\begin{eqnarray}
A_{\text{SIDIS}} (x, Q^2) &=&\frac{\sum_q\, e_q^2\, h_1^q(x, Q^2) n_q^{\uparrow}(Q^2)}{\sum_q\, e_q^2\, f_1^q(x, Q^2) n_q(Q^2)}\enspace.
  \label{eq:asy_dis}
\end{eqnarray}

\begin{figure}[htb]
  \centering
  \includegraphics[width=0.46\textwidth]{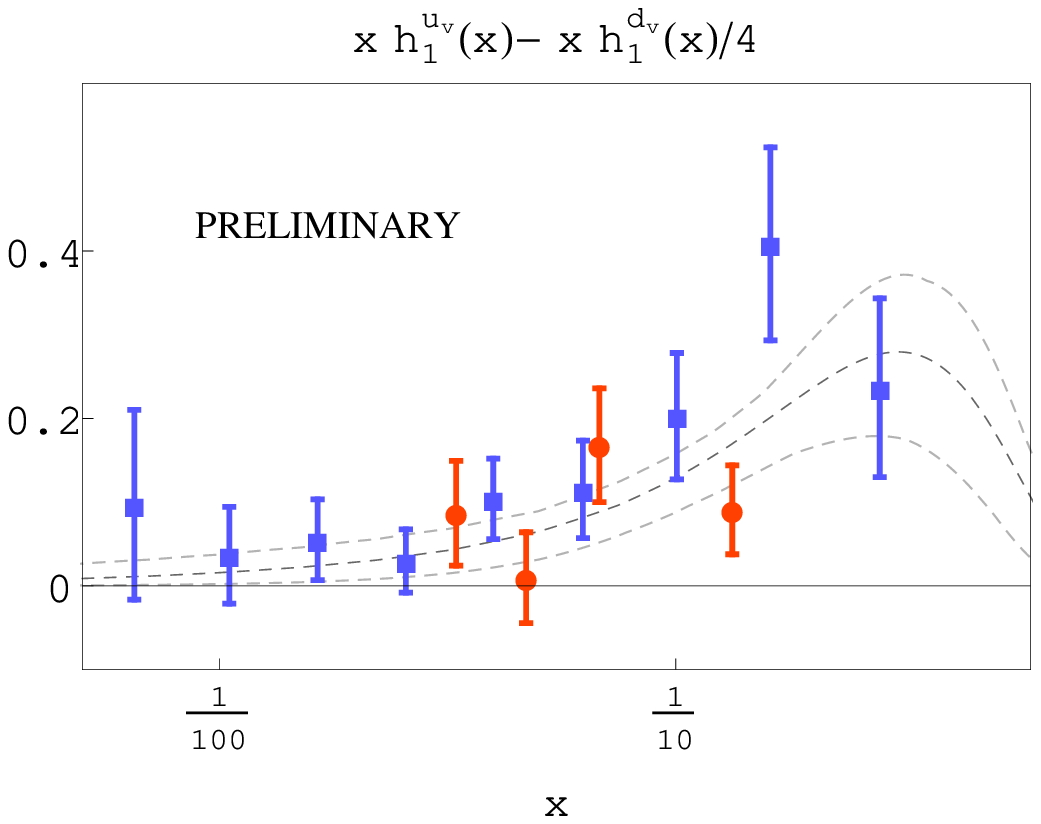}
   \includegraphics[width=0.46\textwidth]{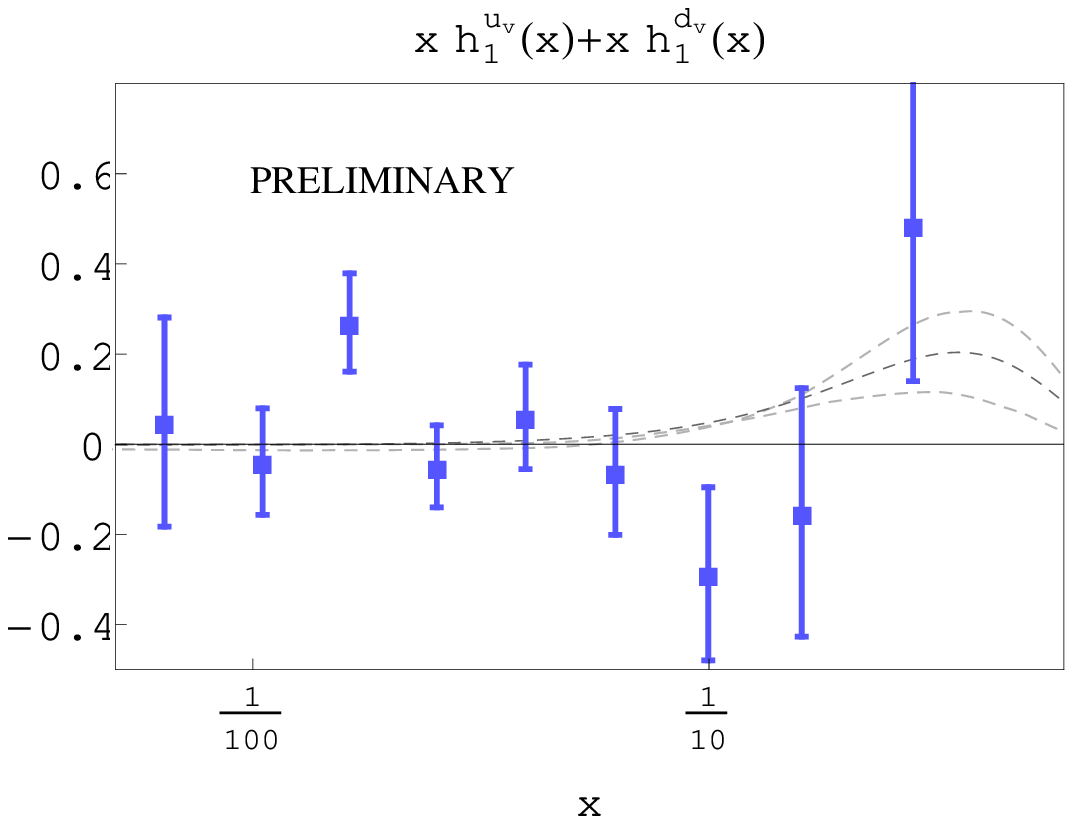}
  \caption{The combinations of Eq.~(\ref{eq:proton}), left panel, and Eq.~(\ref{eq:deuteron}), right panel. The circle red dots are the extracted transversities at HERMES ; the squared blue dots are the extracted transversities at COMPASS. The dashed lines correspond to Torino's transversity~\cite{Anselmino:2008jk}. }
  \label{Fig:proton}
\end{figure}

\begin{figure}[htb]
  \centering
 \includegraphics[width=0.46\textwidth]{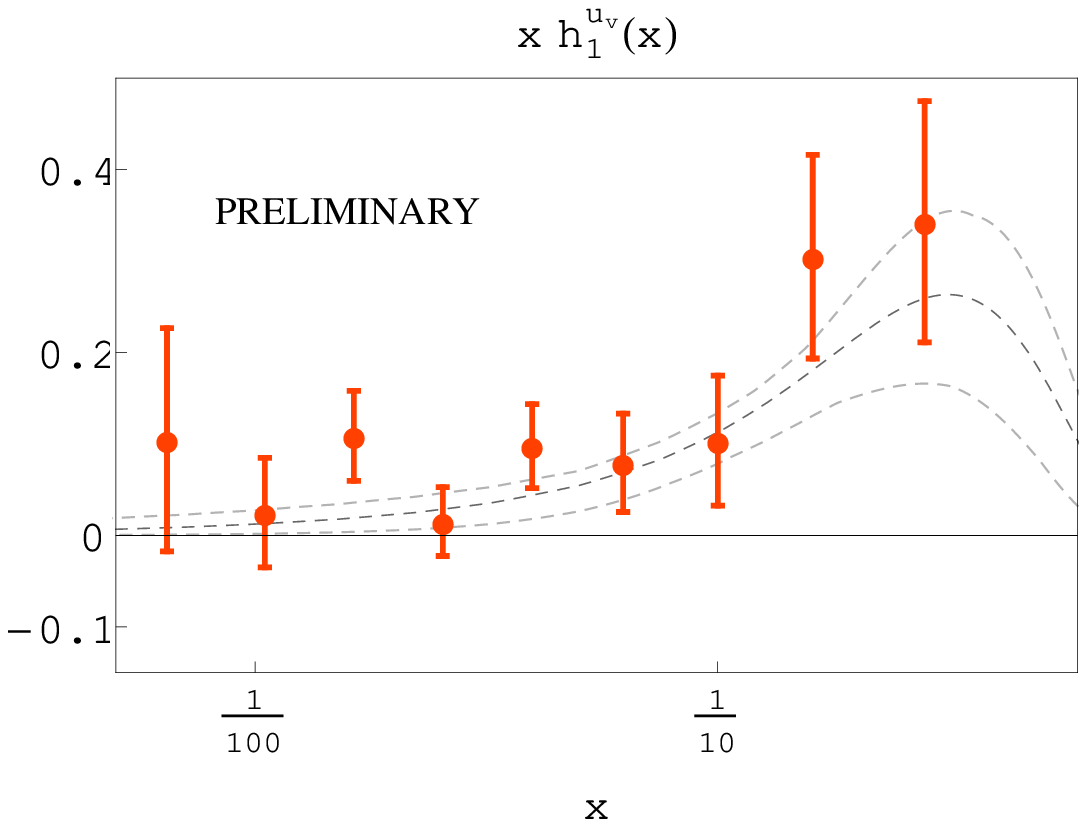}
  \includegraphics[width=0.46\textwidth]{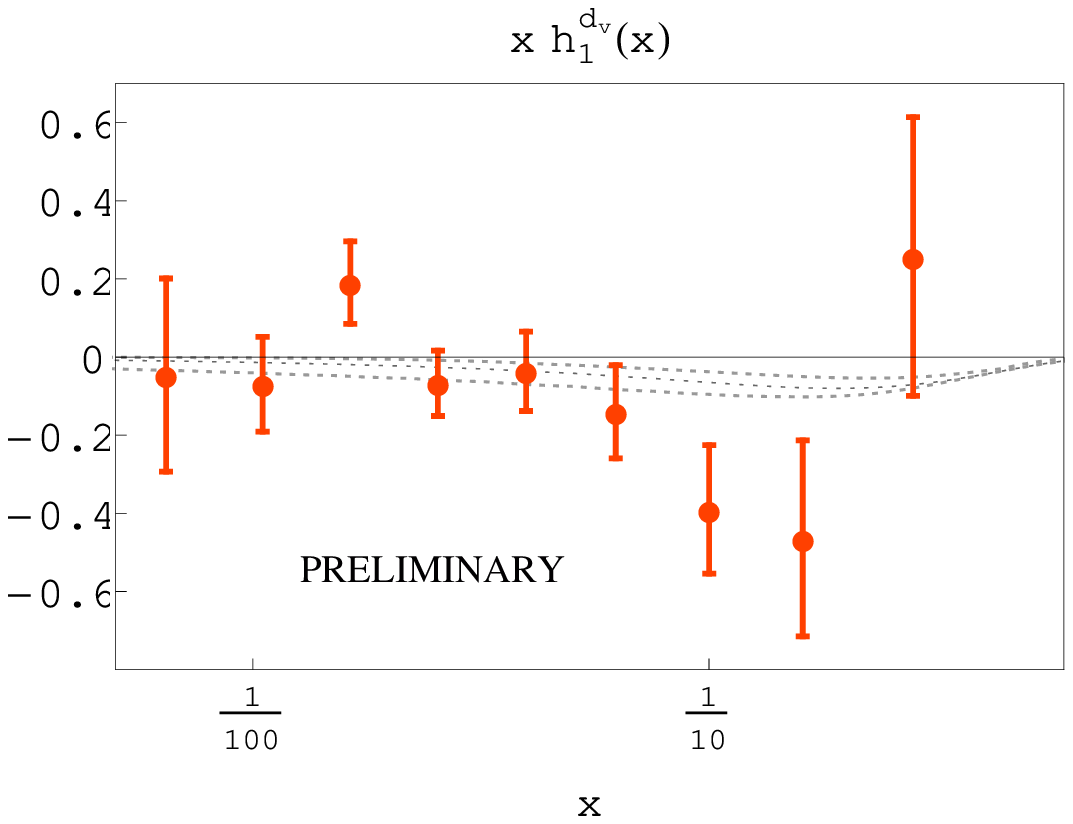}
  \caption{The extracted transversities at COMPASS for the $u$-flavor on the left panel, and for the $d$-flavor on the right panel. }
  \label{Fig:up-down}
\end{figure}

Asumming  isospin symmetry and charge conjugation~\cite{Bacchetta:2006un}, i.e.
\begin{eqnarray}
D_{1}^{u} = D_{1}^{d} = D_{1}^{\bar{u}} = D_{1}^{\bar{d}} \; , \quad D_1^s = D_1^{\bar{s}} \; , \quad D_1^c = D_1^{\bar{c}} \; , \nonumber\\
H_{1}^{\sphericalangle u} = - H_{1}^{\sphericalangle d} = - H_{1}^{\sphericalangle \bar{u}} = 
H_{1}^{\sphericalangle \bar{d}}  \quad \& \quad H_{1}^{\sphericalangle s, \bar s} =H_{1}^{\sphericalangle c, \bar c} =0 \; ,
\end{eqnarray}
the only relevant quantities here are $n_u, n_s$ and $n_u^\uparrow$, since the charm unpolarized PDF can be neglected.
For phenomenological purposes, we evaluate the integrals~(\ref{eq:nD1}, \ref{eq:nH1angle}) over the corresponding kinematical ranges for the HERMES and the COMPASS data. The results are given in Table~\ref{tab:nq_ratios}. Hence, for a proton target, Eq.~(\ref{eq:asy_dis}) leads to the combination of transversity PDFs
\begin{eqnarray}
x h_1^{u_v}(x, Q^2) - {\textstyle \frac{1}{4}}\, x h_1^{d_v}(x, Q^2) 
&=& -  \, \frac{ A^P_{\text{SIDIS}} (x, Q^2)  }{n_u^{\uparrow} (Q^2)} 
\sum_{q=u,d,s} \frac{e_q^2}{e_u^2} \,n_q (Q^2)\, x f_1^{q+\bar{q}}(x, Q^2)\enspace,
  \label{eq:proton}
  \end{eqnarray}
and, for a deuteron target,
\begin{eqnarray}
  x h_1^{u_v}(x, Q^2)+ x h_1^{d_v}(x, Q^2)&=&-  \frac{A^D_{\text{SIDIS}}(x, Q^2)}{n_u^{\uparrow}(Q^2)} \,\frac{5}{3}\, x\left( n_u(Q^2)\,\left(f_1^{u+\bar u}(x, Q^2)+ f_1^{d+\bar d}(x, Q^2)\right)\right.\nonumber\\
&&\hspace{2.3cm}\left.  +\frac{2}{5} n_s(Q^2)\,f_1^{s+\bar s}(x, Q^2)\right)\enspace.
  \label{eq:deuteron}
\end{eqnarray}
In Fig.~\ref{Fig:proton}, we show the extracted transversity combinations in the experimental $x$ values, together with the parametrization of Ref.~\cite{Anselmino:2008jk}. 
We have employed the MSTW08LO PDF 
set~\cite{Martin:2009iq} for the unpolarized PDFs. The errorbands include both the error on the experimental data, $\Delta A_{\text{SIDIS}}$, and the error coming from the fit of $H_1^{\sphericalangle}$, i.e. $\Delta n_u^{\uparrow}$.
Combining the proton and deuteron data, i.e. Eqs.~(\ref{eq:proton}-\ref{eq:deuteron}), allows for a flavor separation, which results are shown in Fig.~\ref{Fig:up-down}.
\\

In summary, we have presented a determination of the transversity parton 
distribution for the $u$ and $d$ flavors in the framework of collinear factorization by using data for pion-pair 
production in DIS off transversely polarized targets, combined with data 
of $e^+ e^-$ annihilations into pion pairs. The final trend of the extracted transversity 
seems not to be in disagreement with the Torino's transversity~\cite{Anselmino:2008jk}, even though there is a large uncertainty due to the deuteron data. More data are needed to clarify the issue. 

\section*{Acknowledgments}

We acknowledge useful discussions with C.~Elia. This work is partially supported by the Italian MIUR through the PRIN 2008EKLACK, and by the Joint Research Activity ``Study of Strongly Interacting Matter" (acronym HadronPhysics3, Grant Agreement No. 283286) under the 7th Framework Programme of the European Community and the Belgian Fund F.R.S.-FNRS via A.~Courtoy's contract of Charg\'ee de recherches.

\bibliographystyle{DISproc}
\bibliography{mybiblio}


\end{document}